\documentclass[letter,
               refpage       
               keeplastbox,   
               nospread,     
               ]{jacow}
\usepackage{pdfpages,multirow,ragged2e} %
\makeatletter%
	\ifboolexpr{bool{xetex}}
	 {\renewcommand{\Gin@extensions}{.pdf,%
	                    .png,.jpg,.bmp,.pict,.tif,.psd,.mac,.sga,.tga,.gif,%
	                    .eps,.ps,%
	                    }}{}
\makeatother

\ifboolexpr{bool{xetex} or bool{luatex}} 
 {}                                      
 {\usepackage[utf8]{inputenc}}           

\usepackage[USenglish]{babel}

\ifboolexpr{bool{jacowbiblatex}}%
 {%
  \addbibresource{jacow-test.bib}
  \addbibresource{biblatex-examples.bib}
 }{}
\usepackage{amsmath}
\usepackage{amssymb}
\usepackage{enumitem}

\begin{document}
\title{Circular Modes for Mitigating Space-Charge effects\\ and enabling Flat Beams\thanks{\NoCaseChange{This work was supported by the U.S. Department of Energy, under Contract No. DE-AC02-06CH11357.}}}

\author{Onur Gilanliogullari\thanks{ogilanli@hawk.iit.edu}, Illinois Institute of Technology, Chicago, IL, USA \\
		Brahim Mustapha, Argonne National Laboratory, Lemont, IL, USA \\
		Pavel Snopok, Illinois Institute of Technology, Chicago, IL, USA}
	
\maketitle

\begin{abstract}
    Flat beams are preferred in high-intensity accelerators and high-energy colliders due to one of the transverse plane emittances being much smaller than the other, which enhances luminosity and beam brightness. However, flat beams are not desirable at low energies due to space charge forces which are significantly enhanced in one plane. The same is true, although to a lesser degree, for non-symmetric elliptical beams. To mitigate this effect and enable flat beams at higher energies, circular mode beam optics can be used. In this paper, we show that circular mode beams offer better control of space charge effects at lower energies and can be transformed into flat beams at higher energies.
\end{abstract}

\section{Introduction}

High-energy colliders and storage rings require high collision luminosity and beam brightness for future scientific discovery and applications. Flat beams could enable this need because one of the transverse beam sizes is much smaller, which enhances luminosity and beam brightness. However, at low energy, flat beams can't sustain high beam currents due to space charge effects which cause tune shifts and unstable motion. To mitigate these effects, flat beams can be propagated as circular mode beams through the lattice while maintaining intrinsic flatness through coupling, then converted to flat beams at high energy. The original idea of circular modes was introduced by Derbenev~\cite{derbenev1998adapting} for an electron cooling experiment at Fermilab. Their theory was further developed by Burov \emph{et al.}~\cite{burov2002circular}. Burov also proposed that circular modes can be used in high-energy lattices to be transported and converted to flat beams for luminosity enhancement~\cite{burov2013circular}. Recently, there has been significant interest in beams with non-zero angular momentum being used to mitigate space charge, either in self-consistent distributions~\cite{hoover2021computation} or in hollow rotating beams~\cite{derbenev2005method}. 

In this work, we have developed different lattice designs for low-energy high-intensity beams that are capable of propagating and maintaining circular mode beams. We will look at circular modes formed by skew triplet transformation (adapter) of Gaussian distributions in periodic lattices and evaluate space charge tune shift performance at high current using two simulation codes WARP~\cite{WARPweb} and TRACK~\cite{TRACKweb}.

\section{Theory}

\begin{figure}
    \centering
     \includegraphics[width=0.49\linewidth]{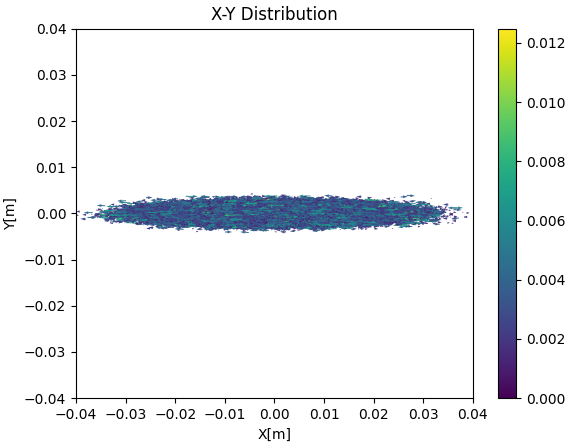}
    \includegraphics[width=0.49\linewidth]{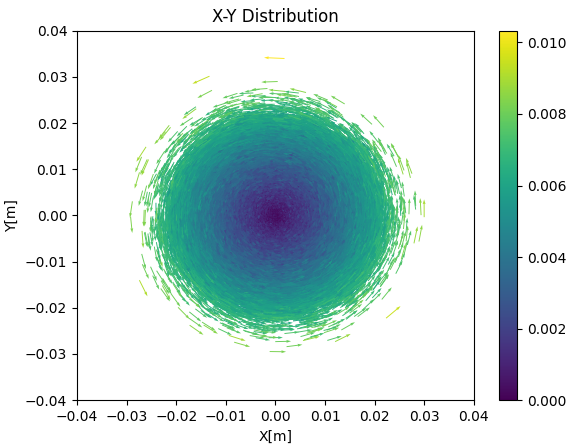}
    \caption{(a) Flat beam, (b) after adapter, circular mode.}
    \label{fig:circular_mode}
\end{figure}

The theory of circular modes is well-understood and discussed in Ref.~\cite{burov2002circular}. There are multiple ways to create circular modes. One of them, as presented in~\cite{derbenev1998adapting, burov2002circular, burov1998insertion}, is to use a skew quadrupole triplet. The skew triplet transforms a flat beam into a round circular-mode beam, as shown in Fig.~\ref{fig:circular_mode}. Here, the color depicts transverse momentum strength, and the arrows show the direction of the particle motion in the transverse plane. We clearly see that transforming a flat beam into a circular mode decreases the beam size in $x$ and increase it in $y$, leading to a round beam with non-zero canonical angular momentum.

The canonical angular momentum of the beam is given by $L_{z}=\epsilon_{\textnormal{I}}-\epsilon_{\textnormal{II}}$~\cite{burov2002circular}, where $\epsilon_{\textnormal{I,II}}$ are the eigenmode emittances, with the skew triplet transforming the initial emittances $\epsilon_{x}\rightarrow\epsilon_{\textnormal{I}}$ and $\epsilon_{y}\rightarrow\epsilon_{\textnormal{II}}$. In Ref.~\cite{burov2002circular}, these eigen-emittances are called $\epsilon_{+}$ and $\epsilon_{-}$ corresponding to the opposite directions of rotation. Because of the coupling, the eigenmode emittances are conserved rather than the $2D$ phase space emittances $\epsilon_{x,y}$. The skew triplet transformation and mapping can be found in~\cite{burov2002circular,burov1998insertion} where the vertical degrees of freedom are coupled to the horizontal degrees of freedom using a vortex condition.

Since a circular mode beam is strongly coupled, the state of the system should be considered in terms
of coupled beam optics using Mais-Ripken parametrization~\cite{willeke1989methods} or its further development by Lebedev \emph{et al.}~\cite{lebedev2010betatron}. The optics can be characterized using the eigenvectors of the system:
\begin{equation}
    \Vec{v}_{\textnormal{I}}=\begin{pmatrix}
    \sqrt{\beta_{x\textnormal{I}}} \\
    -\frac{i(1-u)+\alpha_{x\textnormal{I}}}{\sqrt{\beta_{x\textnormal{I}}}} \\
    \sqrt{\beta_{y\textnormal{I}}}e^{i\nu_\textnormal{I}}\\
    -\frac{iu+\alpha_{y\textnormal{I}}}{\sqrt{\beta_{y\textnormal{I}}}}e^{i\nu_\textnormal{I}}
    \end{pmatrix}, \quad
    \Vec{v}_{\textnormal{II}}=\begin{pmatrix}
    \sqrt{\beta_{x\textnormal{II}}}e^{i\nu_\textnormal{II}}\\
    -\frac{iu+\alpha_{x\textnormal{II}}}{\sqrt{\beta_{x\textnormal{II}}}}e^{i\nu_\textnormal{II}}\\
    \sqrt{\beta_{y\textnormal{II}}}\\
    -\frac{i(1-u)+\alpha_{y\textnormal{II}}}{\sqrt{\beta_{y\textnormal{II}}}}
    \end{pmatrix}.
\end{equation}
Here, $\beta_{jl},\alpha_{jl},u,\nu_{l}$ are the beta functions, alpha functions, coupling parameter, and coupling phases, respectively, with $j~\epsilon~(x,y)$ and $l~\epsilon~(\textnormal{I,II})$. Circular modes produced using an adapter have a special one-to-one transformation because upon calculation of coupled optics functions, one finds that $\beta_{x\textnormal{I}}=\beta_{x\textnormal{II}}=\beta_{y\textnormal{I}}=\beta_{y\textnormal{II}}=\beta_{0}$ and $\alpha_{x\textnormal{I}}=\alpha_{x\textnormal{II}}=\alpha_{y\textnormal{I}}=\alpha_{y\textnormal{II}}=0$. Additionally, there is a circular beta function for which the one-to-one transformation matches the planar beta function to the circular beta function. With this knowledge, it is straightforward to show that $\beta_{0}=\beta_{c}/2$, where $\beta_{c}$ is the circular beta function. The eigenvectors are normalized with respect to the symplectic unit matrix, $\Vec{v}_{\textnormal{I,II}}^{\dagger}S\Vec{v}_{\textnormal{I,II}}=-2i$, where $S$ is the symplectic unit matrix. The phase space vector can be written as:
\begin{equation}
    \Vec{z}=\frac{1}{2}\sqrt{\epsilon_{\textnormal{I}}}\Vec{v_{\textnormal{I}}}e^{i\psi_{\textnormal{I}}} + \frac{1}{2}\sqrt{\epsilon_{\textnormal{II}}}\Vec{v_{\textnormal{II}}}e^{i\psi_{\textnormal{II}}} + C.C.
    \label{eq:phasevector}
\end{equation}

Confining the phase space coordinates to a rotating circle allows the polar coordinate representation: $x=r\cos\theta$ and $y=r\sin\theta$. Applying Courant-Snyder-like formalism~\cite{courant1958theory} will yield the circular optics in the rotating frame with the circular optics function $\beta_{c}$. Implementing the conditions on the phase space vector will result in the vortex condition:
\begin{equation}
    \begin{pmatrix}
    y \\
    y'
    \end{pmatrix} = \begin{pmatrix}
    0 & 1/\beta_{c} \\
    -\beta_{c} & 0
    \end{pmatrix}\begin{pmatrix}
    x \\
    x'
    \end{pmatrix}.
    \label{eq:vortexcondtion}
\end{equation}

To get the same condition as in Eq.~\eqref{eq:vortexcondtion}, the phase space vector can be solved to match the vortex condition. This in return will yield the coupling parameters $u=\frac{1}{2}$ and $\nu_{\textnormal{I,II}}=\pi/2$. With one condition that $\epsilon_{\textnormal{II}}<<\epsilon_{\textnormal{I}}$, which is true for flat beams and hence circular modes, the eigenmode emittances can be found using the normalization condition and phase space vector, $\epsilon_{\textnormal{I,II}}= |\Vec{v}_{\textnormal{I,II}}^{\dagger}S\Vec{z}|^{2}$.
\begin{equation}
    \epsilon_{\textnormal{I,II}}=\frac{x^{2}}{4\beta_{0}} + x'^{2}\beta_{0} + \frac{y^{2}}{4\beta_{0}} + y'^{2}\beta_{0} \pm L_{z},
\end{equation}
where $L_{z}=\epsilon_{\textnormal{I}}-\epsilon_{\textnormal{II}}$. Furthermore, beam sizes can be calculated from the coupled optics functions:
\begin{equation}
    \begin{split}
        \sigma_{x}^{2}&=\epsilon_{\textnormal{I}}\beta_{x\textnormal{I}} + \epsilon_{\textnormal{II}}\beta_{x\textnormal{II}}, \\
        \sigma_{y}^{2}&=\epsilon_{\textnormal{I}}\beta_{y\textnormal{I}} + \epsilon_{\textnormal{II}}\beta_{y\textnormal{II}}.
    \end{split}
\end{equation}

For circular modes, $\epsilon_{\textnormal{II}}<<\epsilon_{\textnormal{I}}$. Here the beam sizes can be written as
\begin{equation}
    \begin{split}
        \sigma_{x}^{2}&=\epsilon_{\textnormal{I}}\beta_{0}(1+\epsilon_{\textnormal{II}}/\epsilon_{\textnormal{I}}), \\
        \sigma_{y}^{2}&=\epsilon_{\textnormal{I}}\beta_{0}(1+\epsilon_{\textnormal{II}}/\epsilon_{\textnormal{I}}),
    \end{split}
    \label{eq:beamsizes}
\end{equation}
where $\sigma_{x,y}$ are the beam sizes. Equation~\eqref{eq:beamsizes} indicates that the beam cross-section is circular. Since the coupling parameter $u=1/2$, which can also be rewritten as $u=\epsilon_{x}/\epsilon_{\textnormal{I}}$, where $\epsilon_{x,y}$ are the apparent emittances, $\epsilon_{x}=\epsilon_{\textnormal{I}}/2$ and $\beta_{0}=\beta_{c}/2$. Therefore, beam sizes can be written as $\sigma_{x,y}^{2}=\beta_{c}\epsilon_{x}$. In other words, a circular mode beam will act as if it is an uncoupled round beam.

\subsection{Angular Momentum Preservation}
Once a circular mode beam is created, it needs to be propagated through the lattice. This can be ensured by preserving the canonical angular momentum (CAM). To preserve CAM, a magnet must have a longitudinal magnetic field, which means a solenoidal field. However, solenoid focusing is not efficient at relatively higher energies and requires very strong fields. Other elements or combinations of elements should be used for beam transport. As shown in Ref.~\cite{burov2002circular}, the most general form of the $4\times4$ transfer matrix that preserves CAM is given by 
\begin{equation}
    \mathcal{M}=R(\theta)\begin{pmatrix}
    T &0 \\
    0 &T
    \end{pmatrix},
\end{equation}
where $R$ is a $4\times4$ rotation matrix with angle $\theta$, and $T$ is a $2\times2$ transverse focusing matrix which is the same in both planes. In other words, to preserve angular momentum, the effective transfer matrix should have symmetric effective focusing in both planes. Among the systems that preserve CAM are the following:
\begin{itemize}[noitemsep]
    \item solenoids,
    \item indexed dipoles ($n=1/2$),
    \item normal quadrupole doublets,
    \item normal quadrupole triplets,
    \item normal quadrupole doublets with normal dipoles,
    \item skew quadrupole doublets.
\end{itemize}

\begin{figure}
    \centering
    \includegraphics[width=0.49\linewidth]{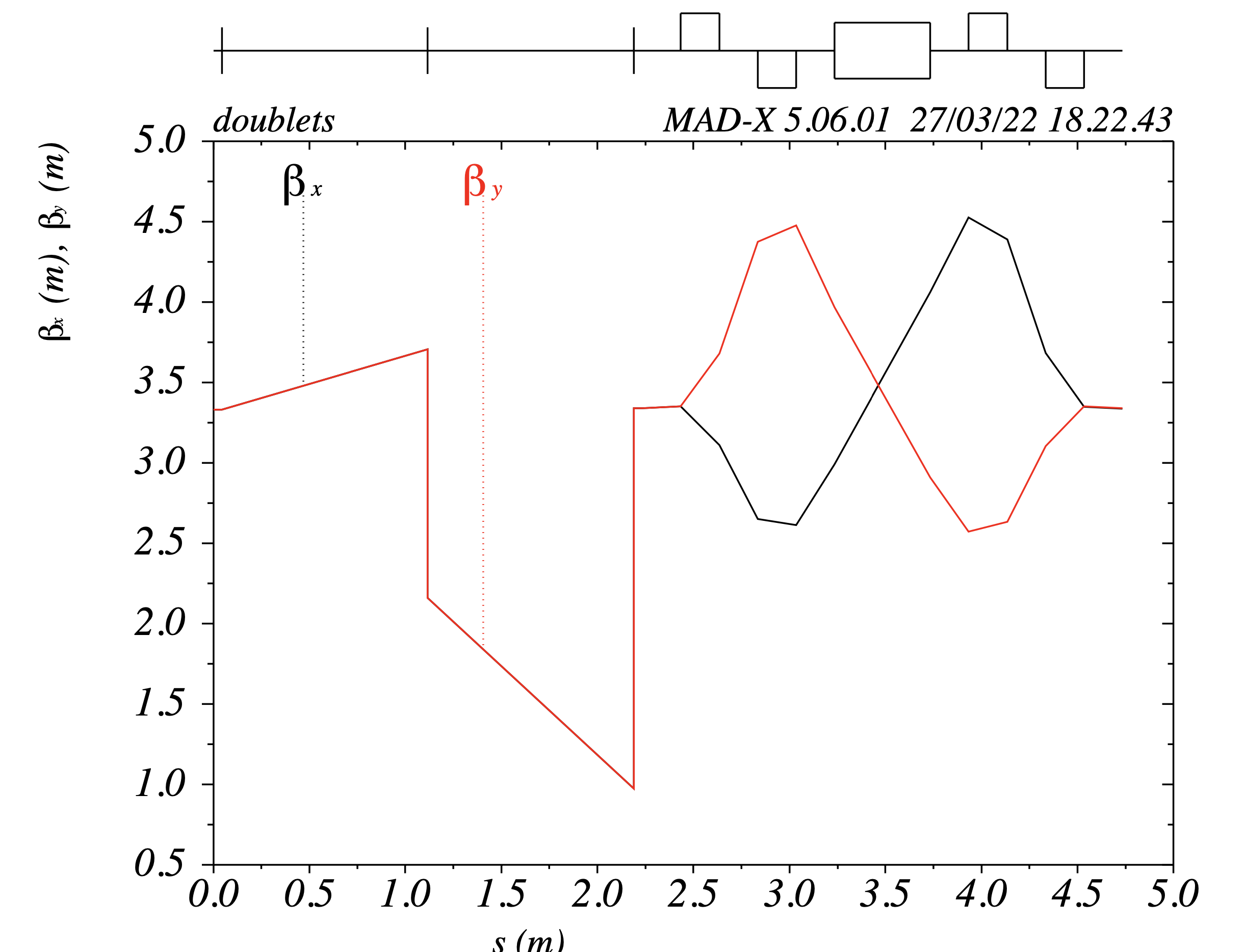}
    \includegraphics[width=0.49\linewidth]{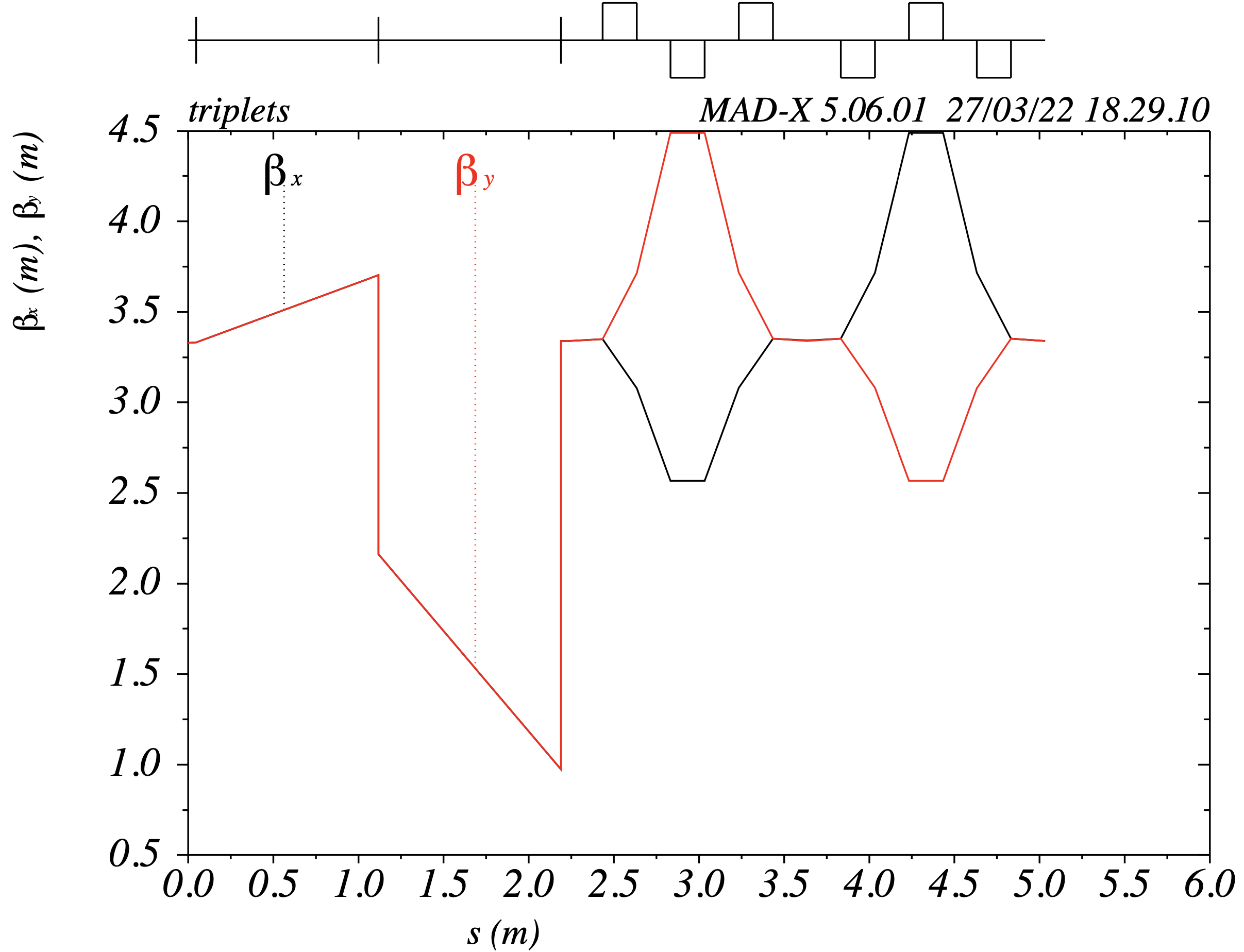}
    \caption{(a) Normal quadrupole doublet with normal dipole, (b) normal quadrupole triplet.}
    \label{fig:quad}
\end{figure}

Normal quadrupole systems can be arranged in a mirror-symmetric fashion to produce equal effective focusing in both planes. The extra condition on the system is that phase advances in both planes should be equal. If phase advances of the system in both planes are not equal, the system does not conserve the circular mode. Mirror-symmetric systems based on normal quadrupoles can be seen in Fig.~\ref{fig:quad}. In this case, the transformation to circular mode is performed using three skew quadrupoles (adapter), then the beam is transported through a normal quadrupole system.

Systems with mirror symmetry break angular momentum at the entrance and restore it at the end since there are no continuous longitudinal magnetic fields. CAM has to be preserved before transforming circular mode back into a flat beam. If systems can not supply equal phase advances and beta functions, the beam will still have rotational motion but will turn into a tilted ellipse (elliptic modes). The advantage of a doublet system with normal dipoles in between is that it allows us to form a ring without the need for indexed dipoles.

\subsection{Space Charge Effect and Circular Modes}

Space charge is mainly an issue at the low-energy stages of acceleration. There are limitations on how much current can be stored due to tune shift and emittance blow-up. The strength of space charge is given by the perveance parameter $\kappa_{SC}$:
\begin{equation}
    \kappa_{\textnormal{SC}}=\frac{eI}{2\pi\epsilon_{0}m_{0}c^{3}\beta^{3}\gamma^{3}},
    \label{eq:kappa}
\end{equation}
where $e$ is the charge of the particle, $I$ the current, $\epsilon_{0}$ the permitivitty of free space, $m_{0}$ the mass of the particle, $c$ the speed of light, $\beta$ and $\gamma$ the relativistic parameters. It is easy to see the dependence on the relativistic parameters as $1/(\beta\gamma)^{3}$, which means the faster we increase the energy the smaller the perveance is. The Coulomb field depends heavily on the size and density of the beam. Since space charge typically has the effect of a defocusing quadrupole, it can be treated as a perturbation. In the case of uncoupled dynamics, space charge tune shift has the form:
\begin{equation}
    \begin{split}
    \Delta Q_{x} &= - \frac{\kappa_{\textnormal{SC}}}{4\pi}\oint\frac{\beta_{x}}{\sigma_{x}(\sigma_{x}+\sigma_{y})}ds, \\ 
    \Delta Q_{y} &= - \frac{\kappa_{\textnormal{SC}}}{4\pi}\oint\frac{\beta_{y}}{\sigma_{y}(\sigma_{x}+\sigma_{y})}ds, \\ 
    \end{split}
    \label{eq:tune_shifts}
\end{equation}
where $\Delta Q_{x}$ is the tune shift in $x$. In the case of coupled dynamics, the theory of tune shift as a perturbation was developed in Ref.~\cite{burov2008coupling}:
\begin{equation*}
    \begin{split}
        \Delta Q_{\textnormal{I}}&= \frac{1}{4\pi}(P_{x}\beta_{x\textnormal{I}} + P_{y}\beta_{y\textnormal{I}}+2P_{xy}\sqrt{\beta_{x\textnormal{I}}\beta_{y\textnormal{I}}}\cos\nu_{\textnormal{I}}), \\ 
    \Delta Q_{\textnormal{II}}&= \frac{1}{4\pi}(P_{x}\beta_{x\textnormal{II}} + P_{y}\beta_{y\textnormal{II}}+2P_{xy}\sqrt{\beta_{x\textnormal{II}}\beta_{y\textnormal{II}}}\cos\nu_{\textnormal{II}}).
    \end{split}
\end{equation*}

Since there is coupling, we talk about modes I and II instead of $x$ and $y$. $P_{x},P_{y}$ and $P_{xy}$ are perturbations in $x$, $y$ or coupled perturbations like skew-quadrupole errors. This equation can be used to calculate the tune shifts for space-charge since the effect is like a defocusing quadrupole perturbation. This allows us to use Eq.~\eqref{eq:kappa}, but the perturbation in $x$ and $y$ depends on the beam distribution, and the tune shift will be similar to Eq.~\eqref{eq:tune_shifts} once it is integrated out for the full system. 

For circular modes, we established the relations with coupled optics parameters, and $P_{xy}$ is zero since it arises from tilted ellipse cross-sections; in addition, a circular mode system has $\nu=\pi/2$, so the cross-terms vanish:
\begin{equation}
        \Delta Q_{\textnormal{I}} = \frac{1}{4\pi}(P_{x} +P_{y})\beta_{0}, \quad \Delta Q_{\textnormal{II}} = \frac{1}{4\pi}(P_{x}+P_{y})\beta_{0}. 
\end{equation}

The perturbations for a Gaussian beam are $P_{x}=\kappa_{\textnormal{SC}}/\sigma_{x}(\sigma_{x}+\sigma_{y})$ and $P_{y}=\kappa_{\textnormal{SC}}/\sigma_{y}(\sigma_{x}+\sigma_{y})$. As shown before, a circular mode beam is round; $\sigma_{x}=\sigma_{y}$ leading to $P_{x}=P_{y}$. Therefore, the tune shift will be the same as for the equivalent round uncoupled beam.

\section{Simulations}

As mentioned above, for the simulations we are using two codes: WARP and TRACK, particle-in-cell (PIC) codes that can effectively compute the space charge effect by solving Poisson's equation. In this section, we will be showing two studies, a normal quadrupole doublet channel and normal quadrupole doublets with normal dipoles in a ring formation. All the systems were designed for a proton beam with $E_{\textnormal{kin}}=10$\,MeV. The beam distributions are all Gaussians with different settings; circular mode beam, uncoupled round beam, and flat beam. Uncoupled round beam and circular mode have the same real plane emittances while the flat beam is the same one that is used for the creation of the circular mode beam. $\epsilon_{x}=10^{-4}$\,m$\cdot$rad and $\epsilon_{y}=10^{-6}$\,m$\cdot$rad for flat with $\beta_{x,y}=3.33$ and $\alpha_{x,y}=0.0$.

\begin{figure}
    \centering
    \includegraphics[width=0.55\linewidth]{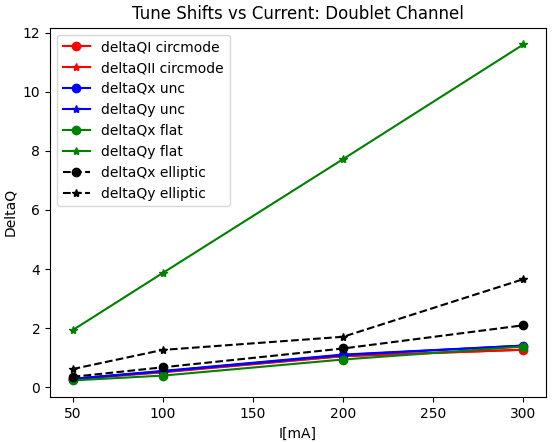}
    \includegraphics[width=0.55\linewidth]{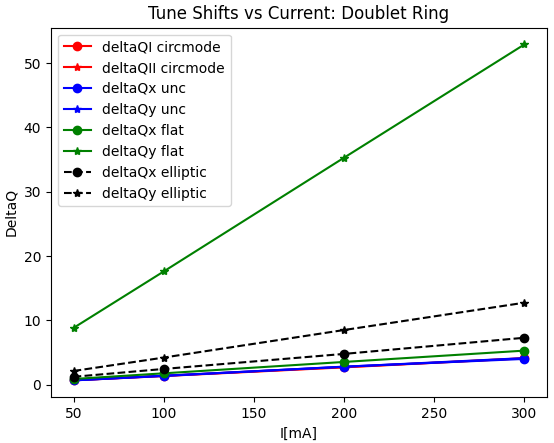}
    \caption{Tune shifts vs. current: (a) quad doublet channel, (b) quad doublet ring.}
    \label{fig:tuneshifts}
\end{figure}

As it can be seen from Fig.~\ref{fig:tuneshifts}, the circular mode tune shifts as predicted above behave the same as for an uncoupled round beam in both cases, whereas for the flat beam, one of the dimensions blows up due to being much smaller. Therefore, circular modes are elegantly solving this problem, as we can see from the results and the analysis in the theory section, where they can be transformed back to flat beams using the inverse transform mentioned in Ref.~\cite{burov2002circular}.

\section{Conclusion}
In conclusion, we showed that circular modes can mitigate the devastating effects of space-charge in flat beams and can be transformed back using the inverse adapter. Tune shifts in flat beams are huge due to the smaller beam size in one dimension, which can promote resonances and beam instability. Circular modes acting like a round beam in real space minimize the tune shift caused by space charge, hence reducing the tune spread. The effectiveness of circular modes can be observed from a practical emittance difference of a factor of $100$, whereas the original theory was derived for one of the emittances being zero.

\section{Acknowledgements}
The authors would like to thank Alexey Burov and Vasiliy Morozov for fruitful discussions on the topic.


\end{document}